\documentclass[prd,aps,twocolumn,showpacs,preprintnumbers,amsmath,amssymb]{revtex4}
\usepackage[dvips]{graphicx} 
\usepackage{epsf}
\usepackage{amsmath} 
\usepackage{amssymb} 
 
\usepackage{graphicx}
\usepackage{dcolumn}
\usepackage{bm}
\def\be{\begin{equation}} 
\def\ee{\end{equation}} 
\def\bea{\begin{eqnarray}} 
\def\eea{\end{eqnarray}}

\begin{document}

\date{\today}

\title{{Non-Gaussianity in a Matter Bounce}}
 
\author{Yi-Fu Cai$^1$\footnote{caiyf@ihep.ac.cn},
Wei Xue$^{2}$\footnote{xuewei@physics.mcgill.ca},
Robert Brandenberger$^{2,1,3}$\footnote{rhb@physics.mcgill.ca},
Xinmin Zhang$^{1,3}$\footnote{xmzhang@ihep.ac.cn}}

\affiliation{1) Institute of High Energy Physics, Chinese Academy of Sciences, P.O. Box 918-4, 
Beijing 100049, P.R. China}

\affiliation{2) Department of Physics, McGill University, 
Montr\'eal, QC, H3A 2T8, Canada}
 
\affiliation{3) Theoretical Physics Center for Science Facilities (TPCSF), Chinese Academy of
Sciences, P.R. China}

\pacs{98.80.Cq}

\begin{abstract}
A nonsingular bouncing cosmology in which the scales of interest today
exit the Hubble radius in a matter-dominated contracting phase
yields an alternative to inflation for producing a scale-invariant spectrum
of adiabatic cosmological fluctuations. In this paper we identify
signatures in the non-Gaussianities of the fluctuations which are
specific to this scenario and allow it to be distinguished from the
results of inflationary models. 

\end{abstract}

\maketitle

\section{Introduction}

One of the problems of the inflationary scenario \cite{Guth}, the current paradigm
of early universe cosmology, is the presence of an initial singularity.
Such a singularity is unavoidable if inflation is obtained from matter
scalar fields in the context of the Einstein action for space-time \cite{Borde}.
As a consequence, there has been a lot of interest in resolving the
singularity by means of quantum gravity effects or effective field
theory techniques. 

A successful resolution of the cosmological singularity may lead to
a non-singular bouncing cosmology. Such non-singular
bounces were proposed a long time ago \cite{Tolman}. They were
studied in models motivated by approaches
to quantum gravity such as the Pre-Big-Bang model \cite{PBB},
higher derivative gravity actions (see e.g. \cite{Markov,nonsing,Tsujik,Biswas}),
brane world scenarios \cite{Varun}
or loop quantum cosmology \cite{LQC}. Nonsingular bounces may
also emerge from non-perturbative superstring theory, as has been
investigated using tools such as $c=1$ matrix theory \cite{Joanna} or
the AdS/CFT correspondence \cite{Sethi,Das}. String gas cosmology
\cite{BV}, an approach to string cosmology in which the temperature
of the universe is non-singular, may also be embedded in a
bouncing cosmology, as realized e.g. in \cite{Biswas2}. Finally, non-singular
bounces may be studied using effective field theory techniques by
introducing a curvature term \cite{Peter} or matter fields violating the 
key energy conditions, for example non-conventional fluids \cite{prequintom,Bozza},
or quintom matter \cite{quintom}. A specific realization of such a quintom
bounce occurs in the Lee-Wick cosmology studied in \cite{LW} \footnote{See
\cite{Novello} for a recent review of bouncing cosmologies.}

In the context of studies of bouncing cosmologies it has been
realized that fluctuations which are generated as quantum vacuum
perturbations and exit the Hubble radius during a matter-dominated
contracting phase lead to a scale-invariant spectrum of cosmological
fluctuations today \cite{FB2,Wands,Wands2,Pinto} (see also \cite{Starob}).
This yields an alternative to cosmological inflation in explaining
the current observational data. We will call this scenario the {\it ``matter
bounce"}. 

It is important to study ways to distinguish the predictions of the {\it
matter bounce} from those of the standard inflationary paradigm.
One criterion is the tensor to scalar ratio which is typically large for a
matter bounce \cite{Wands2,LW}. In this Letter, we study the
distinctive signatures of the {\it matter bounce} in the non-Gaussianities
of the spectrum of cosmological fluctuations. 

We find that the amplitude of the non-Gaussianities is larger than in
simple inflationary models. This is due to the fact that the
fluctuations grow on super-Hubble scales in a contracting universe.
This growth also leads to a specific signature in the shape of
the non-Gaussianities which emerges since it is different solutions
of the fluctuation equation which dominate the spectrum compared
to the case of standard inflation. Both the amplitude and shape of
the three-point function are predicted independent of any free
parameters.

In the following, we first review the key new features which effect
the evolution of cosmological fluctuations in contracting backgrounds,
specifically in the case of the {\it matter bounce}. Then, we turn
to the calculation of the non-Gaussianity function $f_{NL}$ 
and compare our result with the predictions of simple slow-roll
inflation models.

\section{Fluctuations in a Contracting Universe}

To discuss cosmological fluctuations, we write the metric in
the following form (see \cite{MFB} for a detailed exposition
of the theory of cosmological perturbations and \cite{RHBrev1}
for an overview):
\be
ds^2 \, = \, a^2(\eta) \bigl[ (1 + 2 \Phi) d\eta^2 - (1 - 2 \Phi) d{\bf{x}}^2 \bigr] \, ,
\ee
where $\eta$ is conformal time and $\Phi(x, \eta)$ describes the
metric fluctuations \footnote{We are assuming that there is no anisotropic
stress.}. 

To follow fluctuations from the contracting to the expanding phase we will
be  following most of the literature and will use the variable
$\zeta$ which describes the curvature fluctuations in co-moving coordinates.
If we take matter to be a scalar field $\varphi$, then $\zeta$ is
given by 
\be
\zeta \, = \, \bigl( \frac{a}{z} \delta \varphi + \Phi \bigr) \, ,
\ee
where 
\be
z \, = \, a \frac{\varphi'}{\cal{H}} \, ,
\ee
a prime indicating a derivative with respect to conformal time, and ${\cal H}$
standing for the Hubble expansion rate in conformal time.

The variable $\zeta$ is closely related to the Sasaki-Mukhanov
variable \cite{Sasaki,Mukh} in terms of which the
action for cosmological perturbations has canonical kinetic
term:
\be
v \, = \, z \zeta \, ,
\ee
The equation of motion for the Fourier mode $v_k$ of $v$ is
\be \label{EOM}
v_k^{''} + \bigl( k^2 - \frac{z^{''}}{z} \bigr) v_k \, = \, 0 \, .
\ee
If the equation of state of the background is time-independent, then
$z \sim a$ and hence the negative square mass term in (\ref{EOM})
is $H^2$. Thus, on length scales smaller than the Hubble radius,
the solutions of (\ref{EOM}) are oscillating, whereas on larger
scales they are frozen in, and their amplitude
depends on the time evolution of $z$.

On super-Hubble scales, the equation of motion (\ref{EOM})
for $v_k$ in a universe which is contracting or expanding as a 
power $p$ of physical time $t$, i.e.
\be
a(t) \, \sim \, t^p \, , 
\ee 
becomes
\be
v^{''}_k \, = \, \frac{p(2p -1)}{(p -1)^2} \eta^{-2} v_k \, ,
\ee
which has solutions
\be
v(\eta) \, \sim \, \eta^{\alpha}
\ee
with
\be \label{alpha}
\alpha \, = \, \frac{1}{2} \pm \nu \,\,\,\, , \,\,\,\, \nu \, = \, \frac{1}{2} \frac{1 - 3p}{1 - p} \, .
\ee

In the case of an exponentially expanding universe we must take the
$p \rightarrow \infty$ limit of the above solutions. Hence, the solutions
for $\alpha$ are $\alpha = 3/2$ and $\alpha = - 1/2$. The scale
factor $a$ is proportional to $\eta^{-1}$. Hence
\be
\zeta(\eta) \, = \, c_1 \eta^3 + c_2 \, ,
\ee
where $c_1$ and $c_2$ are constants. Since $\eta \rightarrow 0$ as
$t \rightarrow \infty$, it is the constant mode which dominates.
This leads to the well-known result that the fluctuations of $\zeta$
are constant on super-Hubble scales. This also leads to the
conclusion that the power spectrum of $\zeta$ from quantum vacuum
perturbations which are initially on sub-Hubble scales is scale-invariant since
\bea
P_{\zeta}(k, \eta) \, &\equiv& \, \frac{k^3}{12 \pi^2} |\zeta_k(\eta)|^2 \nonumber \\
&\sim& k^3 |\zeta_k(\eta_H(k))|^2 \\
&\sim& k^3 |v_k(\eta_H(k)|^2 a^{-2}(\eta_H(k)) \, \sim \, k^3 k^{-1} k^{-2} \nonumber \\
&\sim& \, {\rm const}  \, , \nonumber 
\eea
making use of the quantum vacuum normalization and the Hubble radius
crossing condition.
 
In the case of a matter-dominated contraction we have $\nu = - 3/2$ and
hence
\be
v_k(\eta) \, = c_1 \eta^2 + c_2 \eta^{-1} \, ,
\ee
where $c_1$ and $c_2$ are again constants. The $c_1$ mode is the 
mode for which $\zeta$ is constant on super-Hubble scales. However,
in a contracting universe it is the $c_2$ mode which dominates and
leads to a scale-invariant spectrum \cite{Wands,FB2,Wands2}:
\bea
P_{\zeta}(k, \eta) \, &\sim& k^3 |v_k(\eta)^2 a^{-2}(\eta) \\
&\sim& \, k^3 |v_k(\eta_H(k))|^2 \bigl( \frac{\eta_H(k)}{\eta} \bigr)^2 \, 
\sim \, k^{3 - 1 - 2} \nonumber \\
&\sim& \, {\rm const}  \, , \nonumber 
\eea
once again using the scaling of the dominant mode of $v_k$, the 
Hubble radius crossing condition $\eta_H(k) \sim k^{-1}$, and
the vacuum spectrum at Hubble radius crossing.

\section{Non-Gaussianities in the Matter Bounce}

\subsection{Formalism}

In this section we will consider non-Gaussianities in the {\it matter bounce}
model. Specifically, we will focus on the amplitude and shape of
the three-point function and on
the function $f_{NL}$ which is commonly used to describe the leading-order
non-Gaussianities.

Non-Gaussianities in single field inflation models were first considered
in \cite{Wise,Hodges,Salopek,Srednicki,Gangui,KomatsuSpergel}  
and in more detail in \cite{Bartolo}
in the context of single field slow-roll inflation models and it was concluded
that the non-Gaussianities would be small. That larger non-Gaussianities
can be obtained in multi-field inflation models or DBI inflation models
was realized in \cite{multifield} and \cite{Tong}, respectively.
An elegant formalism for calculating non-Gaussianities was 
presented in \cite{Maldacena} and extended to the case of generalized
inflation models in \cite{Chen}. For a comprehensive review of
non-Gaussianities the reader is referred to \cite{Komatsu}.

The presence of interactions in the Lagrangian leads to non-Gaussianities.
We will study them following the formalism established in \cite{Maldacena}.
It is easiest to work in the interaction picture, in which the
three-point function to leading order in the interaction coupling
constant is given by
\begin{eqnarray}\label{tpf}
&& <\zeta(t,\vec{k}_1)\zeta(t,\vec{k}_2)\zeta(t,\vec{k}_3)>  \\
&& \,\,\,\, = \, i \int_{t_i}^{t} dt'
<[\zeta(t,\vec{k}_1)\zeta(t,\vec{k}_2)\zeta(t,\vec{k}_3),{L}_{int}(t')]>~,
\nonumber
\end{eqnarray}
where $t_i$ corresponds to the initial time before which
there are any non-Gaussianities. The square parentheses indicate
the commutator, and $L_{int}$ is the interaction Lagrangian (the
integral over space of the interaction Lagrangian density
${\cal L}$ given below).

To calculate the non-Gaussianities in the function $\zeta$, we require the
Lagrangian density for $\zeta$ to cubic order. This has been derived in \cite{Maldacena}
and takes the form
\bea \label{L3}
{\cal L}_3 \, &=& \, 
(\epsilon^2-\frac{\epsilon^3}{2})a^3\zeta\dot\zeta^2 +
\epsilon^2a\zeta(\partial\zeta)^2 -
2\epsilon^2a^3\dot\zeta(\partial\zeta)(\partial\chi) \nonumber \\
&& +
\frac{\epsilon^3}{2}a^3\zeta(\partial_i\partial_j\chi)^2
+ f(\zeta)\frac{\delta{\cal L}_2}{\delta\zeta}|_1 ~,
\eea
where we define $\chi\equiv\partial^{-2}\dot\zeta$ and
$\partial^{-2}$ is the inverse Laplacian. The function $f$ in the last term is
\bea \label{fz}
f(\zeta) \, &=& \, \frac{1}{4{\cal H}^2}(\partial\zeta)^2 -
\frac{1}{4{\cal
H}^2}\partial^{-2}\partial_i\partial_j(\partial_i\zeta\partial_j\zeta) \nonumber\\ 
&& - \frac{a}{{\cal H}}\zeta\dot\zeta 
- \frac{\epsilon a}{2{\cal H}}\partial_i\zeta\partial_i\partial^{-2}\dot\zeta \nonumber \\
&& + \frac{\epsilon a}{2{\cal
H}}\partial^{-2}\partial_i\partial_j(\partial_i\partial^{-2}\dot\zeta\partial_j\zeta)~,
\eea
and $\epsilon$ is given by
\be
\epsilon \, \equiv \, - \frac{{\dot H}}{H^2} \, . 
\ee
In inflationary cosmology, $\epsilon$ is the slow-roll parameter and is
generally much smaller than order unity, whereas in our bouncing
cosmology 
\be
\epsilon \, = \, \frac{3}{2}(1 + w) \, = \, \frac{3}{2}  
\ee
in the contracting phase ($w$ is the equation of state parameter). 

In the standard cosmological model which describes a universe which
was expanding starting from an inflationary phase, $\zeta$ is a constant 
on scales larger than the Hubble radius. On these scales,
the spatial derivatives are negligible, and thus one sees from 
(\ref{fz}) that the interaction Lagrangian vanishes. 
Hence, the integration in (\ref{L3}) runs only up to 
the time of Hubble radius
crossing in the inflationary phase, i.e. one must
just consider the non-linear growth of $\zeta$ inside the Hubble
radius. However, in a contracting universe $\zeta$ grows on
scales larger than the Hubble radius. Hence, the integration
in (\ref{L3}) is dominated by times when the scale is
super-Hubble (until the bounce time, after which $\zeta$
stops growing on super-Hubble scales). 
As we show below, this leads to a 
difference in the shape of the non-Gaussianities. One way to see
this is to note that $\zeta$ is oscillating on scales smaller than the
Hubble radius, whereas the oscillations are frozen out on
super-Hubble scales and the growth of $\zeta$ in the
contracting phase occurs as the increase in the amplitude
of a frozen wave perturbation. The very different time dependence
of $\zeta$ on the scales that dominate the integral (\ref{L3})
leads to a quite different scaling of the integrals, and hence
to a different shape of the non-Gaussianities.

There is another important difference between a bouncing cosmology 
and the inflationary model.  In simple single field slow-roll inflation 
models the cubic Lagrangian is suppressed by
the slow-roll parameter $\epsilon \ll 1$ and so the terms
proportional to $\epsilon^3$ in Eq. (\ref{L3}) can always
be neglected. However, in a bouncing cosmology such as 
the {\it matter bounce} these terms contribute at the same order. 
Moreover, in inflationary cosmology the function $f(\zeta)$ is dominated 
by the first two terms in Eq. (\ref{fz}) because on large scales $\zeta$ is
conserved and thus $\dot\zeta = 0$ for the dominant mode.
In contrast, in the {\it matter bounce} it is the last three terms in Eq. (\ref{fz}) 
which are dominant since the dominant mode of $\zeta$ is
growing as $\eta^{-3}$. Since different terms dominate, we will
get a different shape of the non-Gaussianities.

The three-point function can be expressed in the following
general form:
\bea
<\zeta(\vec{k}_1)\zeta(\vec{k}_2)\zeta(\vec{k}_3)> \, &=& \, 
(2\pi)^7 \delta(\sum\vec{k}_i) \frac{P_{\zeta}^2}{\prod k_i^3} \nonumber \\
&& \times {\cal A}(\vec{k}_1,\vec{k}_2,\vec{k}_3)~,
\eea
where $k_i=|\vec{k}_i|$ and ${\cal A}$ is the shape function whose
amplitude is (in the case of ``local" non-Gaussianities) 
characterized by the non-linearity parameter
$f_{NL}$, where
\be
\zeta \, = \, \zeta_g(x)+\frac{3}{5}f_{NL}\zeta_g^2~,
\ee
and $\zeta_g(x)$ is the linear (and hence Gaussian) part of $\zeta$.
More generally, in momentum space the amplitude ${\cal |B|}_{NL}$
of the non-Gaussianities can be described by
\be
{\cal |B|}_{NL}(\vec{k}_1,\vec{k}_2,\vec{k}_3) \, = \, \frac{10}{3}\frac{{\cal
A}(\vec{k}_1,\vec{k}_2,\vec{k}_3)}{\sum_ik_i^3}~.
\ee

As a last preliminary, we note that in the Lagrangian formalism, 
the curvature perturbation variable $\zeta$ in Fourier
space can be canonically expressed as
\begin{eqnarray}
\zeta(\eta,\vec{k}) \, = \, 
u_k(\eta){a}^\dag_{\vec{k}}+u_k^*(\eta)a_{-\vec{k}}
\end{eqnarray}
with the matter contracting phase mode functions
\begin{eqnarray}
u_k(\eta) \, = \, A
\frac{i[1-ik(\eta-\tilde\eta_B)]}{\sqrt{2k^3}(\eta-\tilde\eta_B)^3}
\exp[ik(\eta-\tilde\eta_B)]~,
\end{eqnarray}
where $\tilde\eta_B$ is the conformal time when the singularity
would occur if the matter contracting phase would continue
to arbitrary densities. The creation and annihilation operators $a$ and $a^\dag$ 
obey the standard canonical commutation relations. The amplitude
$A$ is determined from the quantum vacuum conditions at
Hubble radius crossing in the contracting phase. 
This amplitude determines the power spectrum of $\zeta$.
If we factor out the amplitude and the factor of $k^{-3/2}$, we can define
the following rescaled mode functions:
\begin{eqnarray}
X_k(\eta)\, \equiv \, \frac{1-ik(\eta-\tilde\eta_B)}{(\eta-\tilde\eta_B)^3}e^{ik(\eta-\tilde\eta_B)}~.
\end{eqnarray}

\subsection{Contributions to Non-Gaussianity in the Matter Bounce}

In the following we insert the cubic interaction Lagrangian
(\ref{L3}) into Eq. (\ref{tpf}) and calculate the vacuum
expectation value of the three-point function contributed by the
interaction terms one by one. We evaluate the non-Gaussianity
at the bounce time $\eta_B$.

However, before doing that we employ the same trick as
used in \cite{Maldacena}: In order to cancel the last term in 
Eq. (\ref{L3}), we make the following field redefinition
\be
\zeta \, \rightarrow \, \zeta - f(\zeta)~.
\ee
Inserting this field redefinition into (\ref{tpf}) we find two
terms: the three point function of the rescaled field $\zeta$
on one hand, and terms in which one factor of $\zeta$
has been replaced by $f(\zeta)$ on the other hand.

To compute the first term, we evaluate the right hand
side of (\ref{tpf}) with an interaction Lagrangian which does
not contain the last term in (\ref{L3}). Below, we consider
the contributions of each remaining term in (\ref{L3}).
The second term is called the ``field redefinition term".
It does not involve any integration over time.

Now we consider the individual terms:

\begin{itemize}

\item \textsl{Contribution from the field redefinition}

Since on large scales the first two terms of $f(\zeta)$ can be
neglected, we only need to consider the other three. 
Note that this is precisely the opposite of what happens in
the inflationary paradigm where it is the first two terms
which dominate. As a consequence of this difference, the
field redefinition term leads to a very different shape function.

Moreover, from the solution of the equation of motion
for $\zeta$ one can see that there is an
approximate relation 
\be
\dot\zeta \, \simeq \, -\frac{3}{2}H\zeta
\ee
valid on scales larger than the Hubble radius. Therefore, we obtain the 
following approximate form of the redefinition term in momentum space,
\begin{eqnarray}
\zeta(\vec{k}) \, &\rightarrow& \, \zeta(\vec{k}) -
\int\frac{dk'^3}{(2\pi)^3} \{ \frac{3}{2} +
\frac{9}{8}\frac{\vec{k}'\cdot(\vec{k}-\vec{k}')}{k'^2} \nonumber \\ &-& \, 
\frac{9}{8}\frac{(\vec{k}\cdot\vec{k}')(\vec{k}\cdot(\vec{k}-\vec{k}'))}{k^2k'^2}
\} {\zeta(\vec{k}')\zeta(\vec{k}-\vec{k}')}~, 
\end{eqnarray}
The corresponding shape function is given by (in this and the following
formulas we keep the factor of $\epsilon$ explicit since it will allow us
to understand at which order in $\epsilon$ the key differences in
shape compared to simple slow-roll single field inflation models arise)
\begin{eqnarray}
{\cal A}_{red} \, &=& \, -\frac{\epsilon}{2}\sum k_i^3 -
\frac{\epsilon^2}{32\prod k_i^2}\bigg\{ \sum_{i\neq j}k_i^7k_j^2 +
\sum_{i\neq j}k_i^6k_j^3
\nonumber\\
&& - 2\sum_{i\neq j}k_i^5k_j^4 - 2\sum_{i\neq j\neq k}k_i^5k_j^2k_k^2 - 
\sum_{i\neq j\neq k}k_i^4k_j^3k_k^2 \bigg\} \nonumber \\
&=& \, (-\frac{\epsilon}{2}+\frac{\epsilon^2}{8})\sum k_i^3 +
\frac{\epsilon^2}{32}\sum_{i\neq j}k_ik_j^2   \\
&& - \frac{\epsilon^2}{32\prod k_i^2}\bigg\{ \sum_{i\neq j}k_i^7k_j^2 +
\sum_{i\neq j}k_i^6k_j^3 - 2\sum_{i\neq j}k_i^5k_j^4 \bigg\}~ \, . \nonumber
\end{eqnarray}

Now let us turn to the terms which come from inserting the interaction Lagrangian
into the right-hand side of (\ref{tpf}). These terms all involve an integration over
time from the initial time until the bounce time $\eta_B$. They involve a six point
function of a Gaussian field which yields a cyclic sum of products of three
two-point functions. Thus, the amplitude of the result will be proportional to
the cube of the two point-function. The Lagrangian to be inserted into
(\ref{tpf}) is the integral over space of the Lagrangian density (\ref{L3}).
Each factor of $\zeta$ in (\ref{L3}) is expanded in plane waves. Making
use of
\be
<\zeta^{*}(k) \zeta(k^{'})> \, = \, (2 \pi)^4 k^{-3} \delta^3(k + k^{'}) P_{\zeta}(k) \, ,
\ee
we see that the three momentum integrals are absorbed by the
three delta functions which arise when writing the six-point
function as a product of three two-point functions. The final
integration over space yields an overall factor $\delta^3(\sum_{i = 1}^3 k_i)$
which represents momentum conservation. We will demonstrate
the steps in computing the six-point function for the first contribution,
and simply give the results in the other cases.

\item \textsl{Contribution from the $\zeta\dot\zeta^2$ term}

The contribution of this term in (\ref{L3}) to the three point function is
\begin{eqnarray}
&&(2\pi)^3\delta(\sum\vec{k}_i)\frac{|A|^6}{8\prod
k_i^3}X_{k_1}^*(\eta_B)X_{k_2}^*(\eta_B)X_{k_3}^*(\eta_B)
\\ &&\times
i\int_{-\infty}^{\eta_B}d\eta(\epsilon^2-\frac{\epsilon^3}{2})a^2X_{k_1}X_{k_2}'X_{k_3}'
+ 5 {\rm perms} + c.c. ~, \nonumber
\end{eqnarray}
which gives the following contribution to the shape function
\begin{eqnarray}
{\cal A}_{\zeta\dot\zeta^2} \, = \, (-\frac{\epsilon^2}{12}+\frac{\epsilon^3}{24})\sum
k_i^3~ \, .
\end{eqnarray}

Note that the shape is different from the contribution to the shape function of
the corresponding term in inflationary cosmology, calculated e.g. in \cite{Chen}.
The reason is that in the case of inflation, the time integral runs over times
during which the mode functions are oscillating. Thus, the time integral produces
a factor of $K^{-1}$. In our case, the integral is over super-Hubble scales and
the time integration has a very different result.

\item \textsl{Contribution from the $\dot\zeta\partial\zeta\partial\chi$ term}

A similar calculation shows that the contribution of this term to the shape 
function takes the form
\begin{eqnarray}
{\cal A}_{\dot\zeta\partial\zeta\partial\chi} \, &=& \, 
\frac{\epsilon^2}{24\prod k_i^2}\bigg\{ 2\sum_{i\neq j}k_i^7k_j^2
- 2\sum_{i\neq j}k_i^5k_j^4 \nonumber \\
&& - \sum_{i\neq j\neq k}k_i^5k_j^2k_k^2 \bigg\}  \\
&=& \, -\frac{\epsilon^2}{12}\sum k_i^3 + \frac{\epsilon^2}{12\prod
k_i^2}\bigg\{ \sum_{i\neq j}k_i^7k_j^2 - \sum_{i\neq j}k_i^5k_j^4
\bigg\}~ \, . \nonumber
\end{eqnarray}

Once again, the form is different from that of the contribution of the same
term in inflationary cosmology, for the same reason as explained above.

\item \textsl{Contribution from
the $\zeta(\partial_i\partial_j\chi)^2$ term}

In this case, the contribution to  the shape function is expressed as
\begin{eqnarray}
{\cal A}_{\zeta(\partial_i\partial_j\chi)^2} \, &=& \,
\frac{\epsilon^3}{96\prod k_i^2}\bigg\{ \sum_i k_i^9 -
3\sum_{i\neq j}k_i^7k_j^2  \nonumber \\ && -
\sum_{i\neq j}k_i^6k_j^3 + 3\sum_{i\neq j}k_i^5k_j^4 \nonumber\\
&& - \sum_{i\neq j\neq k}k_i^5k_j^2k_k^2 + \sum_{i\neq j\neq
k}k_i^4k_j^3k_k^2 \bigg\} \nonumber\\
&=& \, -\frac{\epsilon^3}{48}\sum k_i^3 +
\frac{\epsilon^3}{96}\sum_{i\neq j}k_ik_j^2 \\ 
&& + \frac{\epsilon^3}{96\prod k_i^2}\bigg\{ \sum_i k_i^9 -
3\sum_{i\neq j}k_i^7k_j^2 \nonumber \\
&& - \sum_{i\neq j}k_i^6k_j^3 +
3\sum_{i\neq j}k_i^5k_j^4 \bigg\} ~. \nonumber
\end{eqnarray}

\item \textsl{Secondary Contribution}

One may notice that we have neglected the second term of Eq.
(\ref{L3}). Since the form of its shape function is approximately
taken as $(\sum k_i^3)(\sum\frac{k_j^2}{{\cal H}_B^2})$, the
contribution from this term is suppressed on large scales.

\end{itemize}

Finally, summing up all contributions, we obtain
the following shape function of the three-point correlator,
which we separate first into contributions which arise at
various orders in $\epsilon$:
\begin{eqnarray}
{\cal A}^{\epsilon} \, &=& \, -\frac{\epsilon}{2}\sum k_i^3~ \, ,
\end{eqnarray}
\begin{eqnarray}
 {\cal
A}^{\epsilon^2} \, &=& \, -\frac{\epsilon^2}{24}\sum k_i^3 +
\frac{\epsilon^2}{32}\sum_{i\neq j}k_ik_j^2 \\ && +
\frac{\epsilon^2}{96\prod k_i^2}\bigg\{ 5\sum_{i\neq j}k_i^7k_j^2
- 3\sum_{i\neq j}k_i^6k_j^3 -
2\sum_{i\neq j}k_i^5k_j^4 \bigg\}~ \, , \nonumber
\end{eqnarray}
\begin{eqnarray}
{\cal A}^{\epsilon^3} \, &=& \, \frac{\epsilon^3}{48}\sum k_i^3 +
\frac{\epsilon^3}{96}\sum_{i\neq j}k_ik_j^2 \\ && +
\frac{\epsilon^3}{96\prod k_i^2}\bigg\{ \sum_i k_i^9 -
3\sum_{i\neq j}k_i^7k_j^2 \nonumber \\ && - \sum_{i\neq j}k_i^6k_j^3 +
3\sum_{i\neq j}k_i^5k_j^4 \bigg\} ~ \, , \nonumber
\end{eqnarray}
which adds up, for our particular value $\epsilon = 3/2$, to
\begin{eqnarray} \label{result}
{\cal A}_T \, &=& \, \frac{3}{256\prod k_i^2} \bigg\{ 3\sum k_i^9 +
\sum_{i\neq j}k_i^7k_j^2 \nonumber \\
&& - 9 \sum_{i\neq j}k_i^6k_j^3 +5\sum_{i \neq j}k_i^5k_j^4 \\
&& -66 \sum_{i\neq j\neq k}k_i^5k_j^2k_k^2
+9\sum_{i\neq j\neq k}k_i^4k_j^3k_k^2 \bigg\}~. \nonumber
\end{eqnarray}
Comparing with the results for single field slow-roll \cite{Maldacena}
or generalized \cite{Chen} inflationary models, we recognize
some familar terms (the two last terms in (\ref{result})) and some new
terms (the terms in the first two lines of (\ref{result})). The terms
which are different from what is obtained in the case of inflation
arise at second and third order in $\epsilon$. 

Considering the full result (\ref{result}), we see that 
the second to last term has the largest coefficient and hence
dominates. It is the same term which dominates in simple
single-field inflation models. Thus, we conclude that the
dominant term in the non-Gaussianities has local shape,
and an amplitude which is given and independent of any
model parameters. The sign is fixed. 
The new terms which are not present in inflationary cosmology are, however,
not suppressed by more than a factor of order unity. Hence,
with high quality data they could be seen.

\subsection{The amplitude parameter}

There are three forms of non-Gaussianity which are of particular
importance in cosmological observations. They are the ``local form",
the ``equilateral form" and the ``folded form", respectively. In
single field slow-roll inflation models, all three are proportional to slow-roll
parameters and thus are very small. In the matter
bounce, the amplitude of the non-Gaussianities is not 
suppressed by slow-roll parameters. Hence, it is clear that
matter bounces will predict sizable values of these
parameters.

The local form of non-Gaussianity requires that one of the three
momentum modes exits the Hubble radius much earlier
than the other two, for example, $k_1\ll k_2,~k_3$.
Specifically, one is interested in the case when the three momentum vectors
compose an isoceles triangle with $k_1\ll k_2=k_3$. Then one gets
\begin{eqnarray}
{\cal |B|}_{NL}^{\rm local} \, = \, -\frac{35}{8}~,
\end{eqnarray}
which is negative and of order $O(1)$. If our
predicted shape were exactly local (which it is not),
then the above amplitude would equal the famous 
$f_{NL}^{\rm local}$ parameter. Since the matter
bounce model predicts a shape which is loosely
local, one can loosely speaking phrase our
prediction as
\be
f_{NL}^{\rm local} \, = \, -\frac{35}{8}~.
\ee

The equilateral form
requires $k_1= k_2= k_3$. In this case
\begin{eqnarray}
{\cal |B|}_{NL}^{\rm equil} \, = \, -\frac{255}{64}~.
\end{eqnarray}

The folded form of non-Gaussianity with $k_1=2k_2=2k_3$ takes the
value
\begin{eqnarray}
{\cal |B|}_{NL}^{\rm folded} \, = \, -\frac{9}{4}~.
\end{eqnarray}

From the above examples, we see that all of these three 
values of non-Gaussianity are negative and of sizable
amplitude. To quantify this statement, we evaluate the
result numerically 
setting $k_2=k_3=1$ and letting $f_{NL}$ be a function of $k_1$. The
physical value of $k_1$ runs between $0$ and $2$.

\begin{figure}[htbp]
\includegraphics[scale=0.8]{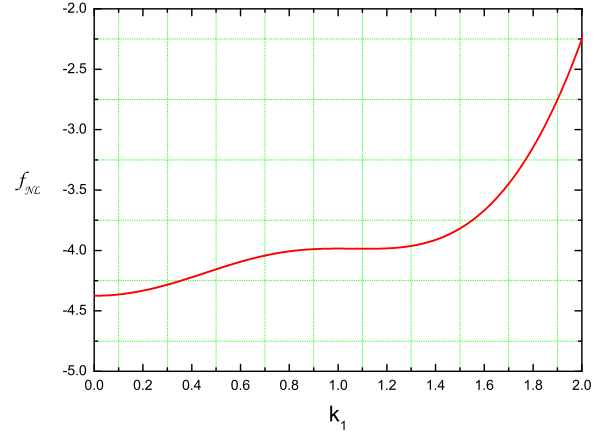}
\caption{The $f_{NL}$ parameter in the Matter Bounce. }
\label{Fig:fnl}
\end{figure}

\subsection{The Shape of Non-Gaussianity in the Matter Bounce}

\begin{figure}[htbp]
\includegraphics[scale=0.7]{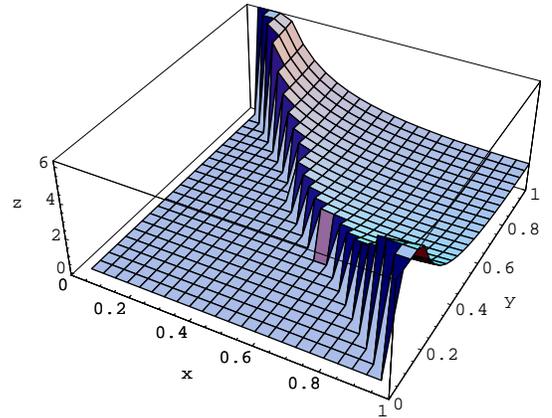}
\caption{The shape of non-Gaussianity in the Matter Bounce. The vertical axis
is the negative of the shape function. This figure shows the contribution
of the terms of order $\epsilon$.}
\label{Fig:epsilon1}
\end{figure}

\begin{figure}[htbp]
\includegraphics[scale=0.7]{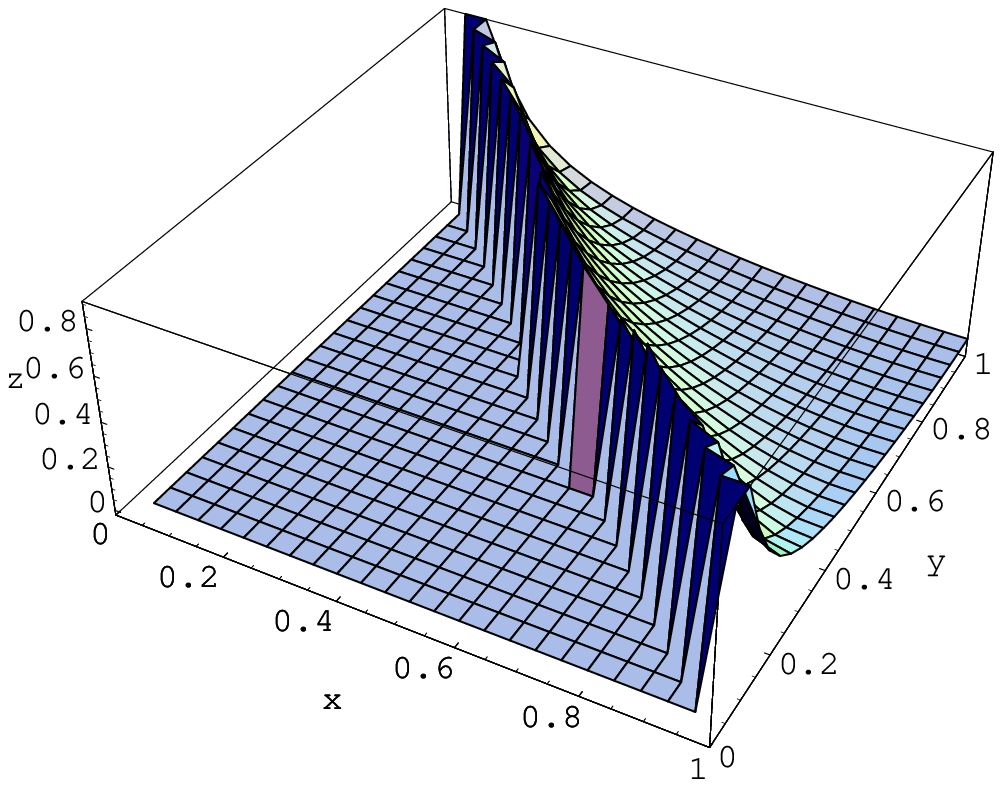}
\caption{The shape of non-Gaussianity in the Matter Bounce. 
This figure shows the contribution
of the terms of order $\epsilon^2$.}
\label{Fig:epsilon2}
\end{figure}

\begin{figure}[htbp]
\includegraphics[scale=0.7]{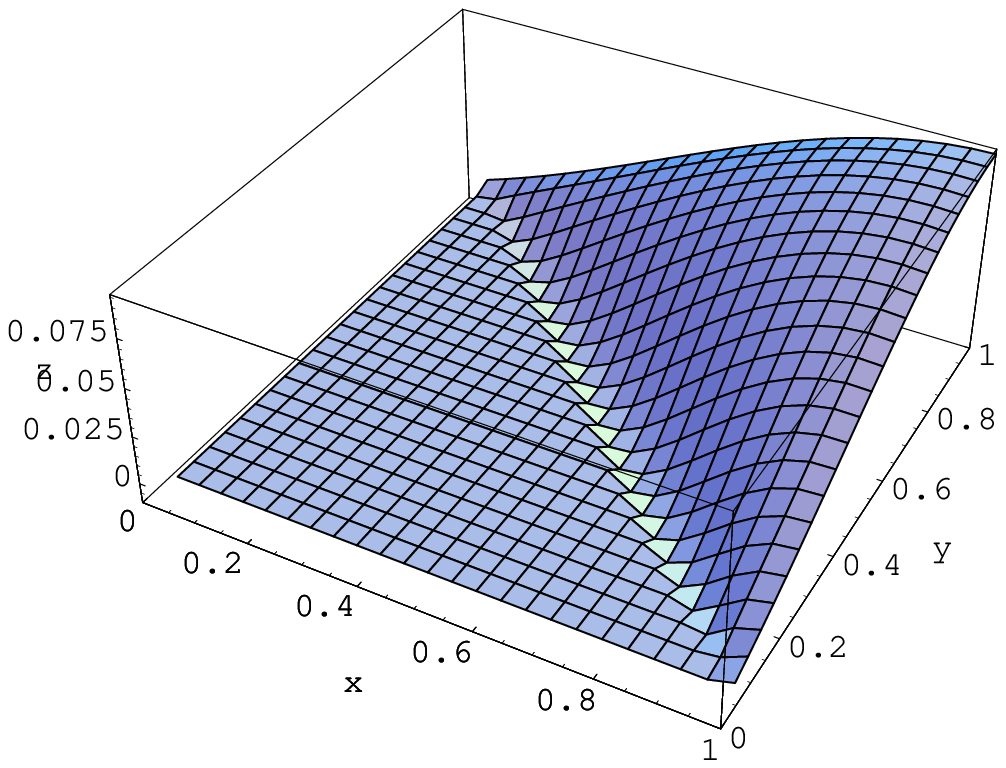}
\caption{The shape of non-Gaussianity in the Matter Bounce. 
This figure shows the contribution
of the terms of order $\epsilon^3$.}
\label{Fig:epsilon3}
\end{figure}

\begin{figure}[htbp]
\includegraphics[scale=0.7]{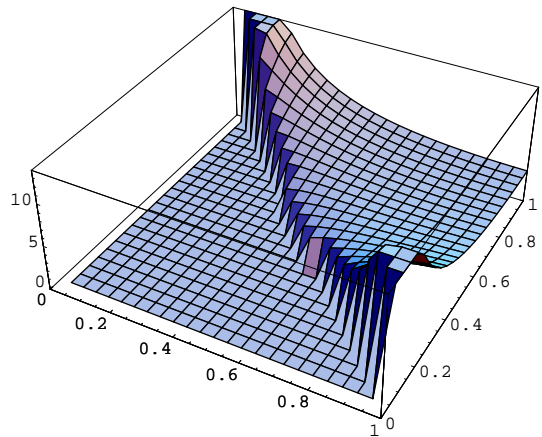}
\caption{The shape of non-Gaussianity in the Matter Bounce. 
This figure shows the contribution of all terms.}
\label{Fig:shape}
\end{figure}

It is interesting to determine the shape of the non-Gaussianities, 
which has the potential to distinguish different cosmological models 
once the data will be sufficiently accurate. 
A useful description of the shape is given by
\begin{eqnarray}\label{shape}
\frac{{\cal A}_T}{k_1k_2k_3}~.
\end{eqnarray}
To obtain a better idea of the shape of the non-Gaussianities, we have
evaluated Eq. (\ref{shape}) numerically. Our results are plotted
in Figures \ref{Fig:epsilon1}), \ref{Fig:epsilon2}), \ref{Fig:epsilon3}) and
\ref{Fig:shape}. In the figures, we use the following
convention: $k_3=1$, the x-axis is $k_1$, the y-axis is $k_2$, and
the z-axis corresponds to the shape ${\cal A}/k_1k_2k_3$. 
Figures \ref{Fig:epsilon1}), \ref{Fig:epsilon2}) and \ref{Fig:epsilon3})
depict the shape functions of the contributions to order $\epsilon$,
$\epsilon^2$ and $\epsilon^3$, respectively, Figure (\ref{Fig:shape})
shows the shape of the total contribution. 

\begin{figure}[htbp]
\includegraphics[scale=0.7]{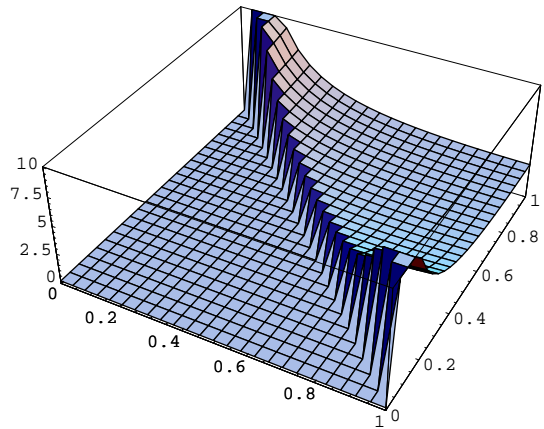}
\caption{The shape of non-Gaussianity in single field
slow-roll inflation to leading order in $\epsilon$. The vertical
axis is the amplitude of the shape function.}
\label{Fig:inflation}
\end{figure}

For comparison, the shape function of the non-Gaussianities in
single field slow-roll inflation to leading order in the slow-roll
parameter $\epsilon$ is shown in Figure \ref{Fig:inflation}.
We see that the dominant structure of the shape function
(modulo sign) is the same in the matter bounce. However,
it is also clear that the sub-leading correction terms in the
case of the matter bounce shape function are clearly visible.
They arise in the terms which are of the order $\epsilon^2$ and
$\epsilon^3$.
Therefore, these next-to-leading correction terms in the case
of slow-roll inflation are suppressed by orders of magnitude
(namely by $\epsilon$) compared to the leading term.
 
From the above analysis, we have learned that
the amplitude of the non-Gaussianities of metric perturbations
predicted in the {\it matter bounce} scenario is of order ${\cal O}(1)$,
much larger than in single-field slow-roll inflation models. The
shape is dominated by a term of local form, but there are
sizable corrections with which the matter bounce model could
in principle be distinguished from large classes of
inflationary scenarios, including many generalized inflation models
\cite{Chen}.

There are two basic reasons leading to these differences. One
is that in a bounce model the analog of the slow-roll parameter
is large, the other is that the
perturbations outside the Hubble radius are not conserved
which provides a new origin to generate non-Gaussianities.

\subsection{Squeezed Limit}

It is also interesting to consider the behavior of shape
function in the squeezed limit when  $k_1= k_2 \equiv k$ and
$k_3\rightarrow0$ \footnote{We thank X. Chen for
encouraging us to consider this limit.}

In this limit, the leading terms of the above three shape functions
for the contributions to orders $\epsilon$, $\epsilon^2$ and $\epsilon^3$
are all proportional to
\begin{eqnarray}
{\cal A} \, \sim \, \frac{k^5}{k_3^2}~ \, .
\end{eqnarray}
However, when we sum the three contributions, we find that the 
leading terms  cancel and the total shape function takes the form
\begin{eqnarray}
{\cal A}_T|_{\rm squeezed} \, = \, -\frac{21}{8}k^3~.
\end{eqnarray}

\section{Discussion and Conclusions}

We have calculated the amplitude and shape of
the non-Gaussianities in the matter bounce model
as quantified by the three-point correlation function
of $\zeta$. Since in this model the fluctuations grow
on super-Hubble scales during the contracting phase,
different terms in the interaction Lagrangian dominate
the contribution to the non-Gaussianities. In addition,
there is no slow-roll suppression of the amplitude.
Hence, both the amplitude and the shape are 
different from what is predicted in inflationary models.

The amplitude  of the non-Gaussianity is 
of the order $O(1)$. Both the amplitude and the
sign are fixed, independent of any model parameters,
different from the result obtained in the standard paradigm. 
The amplitude which we predict is smaller than
the current upper limits from the WMAP results
\cite{WMAP}, but they are in the range of what
the Planck satellite will be able to observe \cite{Planck}
\footnote{Note that since the shape function
is different from the ones from which the limits
on non-Gaussianities are derived, the WMAP
bounds and Planck anticipated limits cannot be
blindly applied - we thank the participants of the
KITPC Program on ``Connecting Fundamental Theory
with Cosmological Observations'' for stressing
this point to us.}. The
shape function contains contributions which are
different in shape compared to what is obtained in
inflationary models, although the dominant term
is of local form.
These distinctive contributions arise at order $\epsilon^2$
and $\epsilon^3$ in the parameter which in the
case of single field slow-roll inflation is the slow-roll
parameter which is much smaller than $1$, but which
in our case is $3/2$. Hence, in our case the extra
terms in the shape functions
are not suppressed by more than a factor of order
unity compared to the contributions which are of
similar shape to those which dominate in the inflationary
paradigm.

Thus, we conclude that the amplitude and the shape of 
non-Gaussianities of the three-point function provide distinctive
signatures for cosmological models based on a matter
bounce. Since the differences in the shape of the 
non-Gaussianities are mainly due to the growth of
the curvature fluctuations on super-Hubble scales
in the contracting phase, the shape function of
non-Gaussianities in other scenarios in which
adiabatic fluctuations are produced in the
contracting phase will be similar to what we have
obtained here. 

Note that we have considered only the case
of a canonical kinetic term for the matter fields, and
we have focused on the adiabatic mode. Including
non-canonical matter fields and entropy modes
will give further contributions to the non-Gaussianities
which will make the amplitude larger in general,
as the inclusion of such effects increases the amplitude
of non-Gaussianities in the case of inflation models.

The formalism we have developed here can be used
to compute the non-Gaussianities in other collapsing
scenarios, for example the Ekpyrotic model \cite{KOST}.
There has been a lot of recent work on non-Gaussian
fluctuations in the multi-field variant of the scenario
\cite{Ekpyrotic}. However, the adiabatic mode will also
contribute to the non-Gaussianities, and this contribution
can be computed easily using our methods.

\section*{Acknowledgments}

We would like to thank Xingang Chen and
Yi Wang for many helpful discussions. RB wishes to thank the
Theory Division of the Institute of High Energy Physics (IHEP) for
their wonderful hospitality and financial support. RB is also
supported by an NSERC Discovery Grant and by the Canada Research
Chairs Program. The research of Y.C. and X.Z. is supported in part
by the National Science Foundation of China under Grants No.
10533010, 10675136 and 10821063, by the 973 program No. 2007CB815401, and by
the Chinese Academy of Sciences under Grant No. KJCX3-SYW-N2.
One of us (RB) acknowledges hospitality of the KITPC during the
program ``Connecting Fundamental Physics with Cosmological
Observations" while this manuscript was being finalized.

\end{document}